# Dimensional crossover and topological nature of the thin films of a three-dimensional topological insulator by band gap engineering

Zhenyu Wang[1,2,6†*], Tong Zhou[1†], Tian Jiang[1,5†], Hongyi Sun[4,8], Yunyi Zang[3], Yan Gong[3], Jianghua Zhang[3,6], Mingyu Tong[5], Xiangnan Xie[1], Qihang Liu[4,7*], Chaoyu Chen[4*], Ke He[3,6] & Qi-Kun Xue[3,6]

[1]*State Key Laboratory of High Performance Computing, College of Computer, National University of Defense Technology, Changsha 410073, P. R. China*
[2]*National Innovation Institute of Defense Technology, Academy of Military Sciences PLA China, Beijing 100010, P. R. China*
[3]*State Key Laboratory of Low Dimensional Quantum Physics, Department of Physics, Tsinghua University, Beijing 100084, P. R. China*
[4]*Shenzhen Institute for Quantum Science and Engineering (SIQSE) and Department of Physics, Southern University of Science and Technology (SUSTech), Shenzhen 518055, P. R.China*
[5]*College of Advanced Interdisciplinary Studies, National University of Defense Technology, Changsha 410073, P. R. China*
[6]*Beijing Academy of Quantum Information Sciences, Beijing 100084, P. R. China*
[7]*Center for Quantum Computing, Peng Cheng Laboratory, Shenzhen 518055, P. R. China*
[8]*School of Physics, Southeast University, Nanjing 211189, China*

Identification and control of topological phases in topological thin films offer great opportunity for fundamental research and the fabrication of topology-based devices. Here, combining molecular beam epitaxy, angle-resolved photoemission spectroscopy and *ab-initio* calculations, we investigate the electronic structure evolution in $(Bi_{1-x}In_x)_2Se_3$ films ($0 \leq x \leq 1$) with thickness from 2 to 13 quintuple layers. We identify several phases with their characteristic topological nature and evolution between them, *i.e.*, dimensional crossover from a three-dimensional topological insulator with gapless surface state to its two-dimensional counterpart with gapped surface state, and topological phase transition from topological insulator to a normal semiconductor with increasing In concentration x. Furthermore, by introducing In alloying as an external knob of band gap engineering, we experimentally demonstrated the trivial topological nature of $Bi_2Se_3$ thin films (below 6 quintuple layers) as two-dimensional gapped systems, in consistent with our theoretical calculations. Our results provide not only a comprehensive phase diagram of $(Bi_{1-x}In_x)_2Se_3$ and a route to control its phase evolution, but also a practical way to experimentally determine the topological properties of an gapped compound by topological phase transition and band gap engineering.

One of the central goals of modern condensed matter physics is to realize novel quantum phases of matter and develop practical control over their properties, holding promise for new generation of photonic, electronic and energy technologies. Topological insulators (TIs) are new phases of quantum matter with unique bulk band topology and surface states protected by certain symmetries.[1-3] For example, the strong TIs,[4] classified by a topological $Z_2$ invariant $v_0 = 1$, are protected by time-reversal symmetry and thus possess metallic surface states at the boundary to a trivial insulator[5,6] (*e.g.*, band insulator or vacuum, $v_0 = 0$) with an odd number of Dirac cones. The discovery of the TIs has led to the prediction and realization of other new topological states including topological crystalline insulator (TCI),[3] topological Kondo insulator.[7] topological Dirac/Weyl semimetal[8] and so on. The emergency of such rich topological systems offers a fertile ground, for not only the exploration of exotic quantum phenomena such as supersymmetry state,[9] interacting TIs,[10] Floquet topological insulator[11] and so on, but also the realization of novel topology-based devices like topological transistor,[12] spin torque device[13] and topological photodetectors.[14]

For most of such applications, ultrathin films (few atomic layers thick) of three-dimensional (3D) TIs with well-controlled number of layers and composition are inevitably required. During the truncation along one direction, the hybridization between opposite surfaces opens a gap in the surface Dirac cone and drives the system into a phase with gapped surface states. For a typical 3D TI such as $Bi_2Se_3$, an interesting question thus arises: if the thin films with observable hybridization gap,[15] i.e., 1-5 quintuple layers (QLs, where 1 QL ≈ 1 nm), manifest two-dimensional (2D) quantum spin Hall insulators with an inverted band gap or not? It was theoretically predicted by continuous Hamiltonian model and density functional theory (DFT) calculations that the topological feature of $Bi_2Se_3$ thin films exhibits an oscillatory manner with increasing thickness.[16,17] Nevertheless, angle-resolved photoemission spectroscopy (ARPES) measurements directly performed on $Bi_2Se_3$ thin films cannot tell the sign of the hybridization gap,[15] while transport experiments have to suffer the bulk conductivity because of the intrinsic n-type doping.[18,19] Therefore, no conclusive answer to the question has been drawn yet.

To understand the topological nature of a new quantum phase, a powerful approach is to study its evolution from an understood phase through phase transition. Topological phase transition (TPT) refers to transition from one topologically nontrivial phase to another or to a topologically trivial phase, and vice versa. The realization of TPT involves utilization of structural or compositional degrees of freedom as an external "knob", such as alloying a normal insulator (NI) with TI components or applying strain or electric field to a NI, attempting to induce thereby band inversion.

For example, BiTlSe$_2$ bulk crystal doped with S, BiTl(S$_{1-x}$Se$_x$)$_2$, is the first system where TPT was observed.[20,21] TPT from topological-metallic to normal-insulating states has further been realized in (Bi$_{1-x}$In$_x$)$_2$Se$_3$[19,22-24] and (Bi$_{1-x}$Sb$_x$)$_2$Se$_3$[25] single crystals and films. To date, the experimental exploration of TPT has been mainly limited on thick films and bulk samples,[20-22,24-26] whereas only few thickness-depended transport measurements have focused on ultrathin (Bi$_{1-x}$In$_x$)$_2$Se$_3$ films down to 2 QLs.[19,23] For ultrathin (Bi$_{1-x}$In$_x$)$_2$Se$_3$ samples, with increasing In substitution, the spin-orbit coupling (SOC) strength decreases and drives the system from a topologically nontrivial state to a trivial state, accompanied by the inverted bulk gap closure and reopening.[19,23] During such process, hybridization comes into play due to the increase of the surface penetration depth with decreasing bulk gap size, manifesting itself as a dimensional crossover as the films change from being 3D to approximately 2D. For ultrathin TI films, this picture has been merely deduced from transport behaviors,[19,23] while lacks direct and detailed band structure investigation [say, via ARPES]. Furthermore, the general picture of TPT in (Bi$_{1-x}$In$_x$)$_2$Se$_3$ still contains controversial description on the band structure evolution concerning whether it is a linear gap-closure[19,22,23] or a sudden gap-closure scenario.[24]

In this work, we study the topology and TPT of ultrathin Bi$_2$Se$_3$ films by employing structural and compositional "knobs". Using molecular beam epitaxy (MBE) we have grown high-quality indium doped Bi$_2$Se$_3$ films ((Bi$_{1-x}$In$_x$)$_2$Se$_3$) with thickness ranging from 2 QLs to 13 QLs, and In substitution x from 0 to 1. Systematic *in situ* ARPES measurements have been performed to monitor the evolution of both bulk and surface band structure. A dimension (thickness)-substitution phase diagram has been drawn. In this phase diagram, we have identified several different quantum states, *i.e.*, 3D TI with gapless surface states and its 2D counterpart hosting hybridization-induced gapped surface states (hereafter referred to as 2D GSS) below the critical substitution concentration (x≤5%), and NI state (9%≤x≤100%). The dimensional crossover from 3D TI to 2D GSS is driven by both SOC and surface state hybridization, and thus can be manipulated through bulk band gap engineering by either thickness control or substitution control, while the transition from 3D TI or 2D GSS to NI is driven solely by SOC, independent of film thickness. More importantly, we illustrate that one can create a phase evolution process by introducing an external knob and monitor the existence of TPT in order to determine the topological nature of the initial phase. Our ARPES-measured gap evolution of thin film $(Bi_{1-x}In_x)_2Se_3$ (e.g., 4 QLs) by tuning the amount of In alloying shows a monotonic increasing feature, indicating that as 2D systems, the 2D GSS films, particularly $Bi_2Se_3$ ultrathin films with an observable hybridization gap (i.e., 1-5 QLs) are all topologically trivial. Such results are confirmed by our DFT calculated $Z_2$ invariants by using hybrid correlation-exchange functional. Our findings provide a practical way to control the dimensional crossover of the topological phase evolution in $(Bi_{1-x}In_x)_2Se_3$ ultrathin films by composition alloying, which is critical for potential device application. Furthermore, we provide an effective way to experimentally determine the topological nature of an insulating phase by monitoring the band gap evolution under the tuning of an external knob, such as SOC and tensile strain.

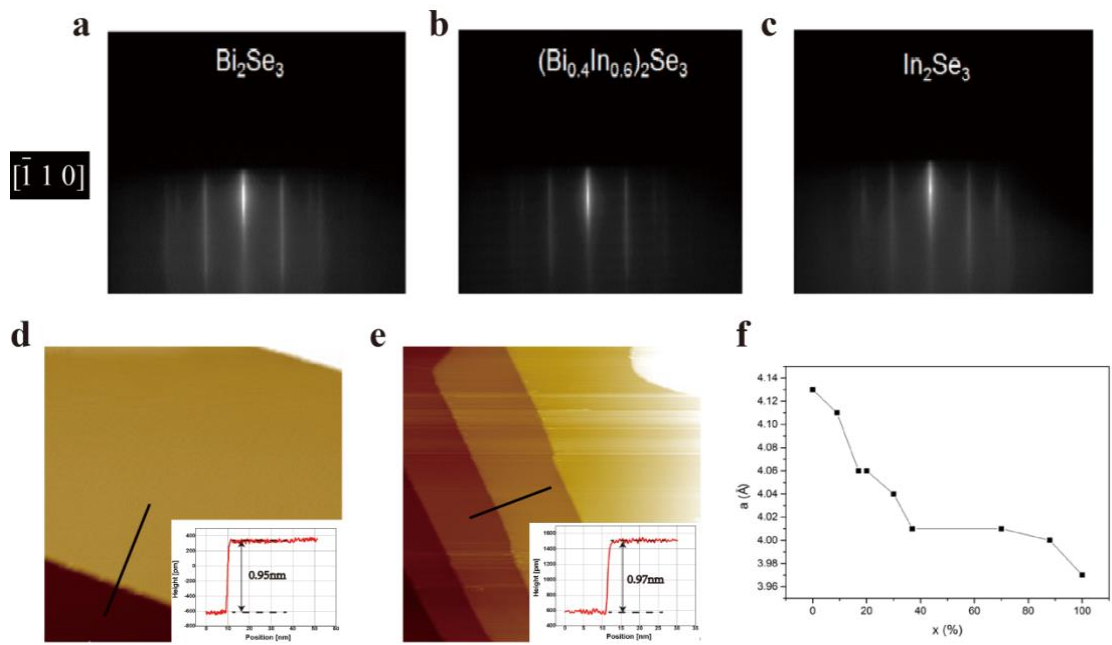

**Figure 1. Characteristics of (Bi$_{1-x}$In$_x$)$_2$Se$_3$ thin films by RHEED and STM. (a,b,c)**, Bi$_2$Se$_3$, (Bi$_{0.4}$In$_{0.6}$)$_2$Se$_3$ and In$_2$Se$_3$ RHEED patterns. **(d),** STM image of Bi$_2$Se$_3$ film (4 QLs, 65 nm × 65 nm) on graphene substrate. The inset shows STM line profile along the black lines, which shows a 0.95 nm step corresponding to the height of 1 QL. **(e),** STM image of In$_2$Se$_3$ film (13 QLs, 100 nm × 100 nm) on graphene substrate. The inset also shows the corresponding line profile. **(f),** Substitution x dependent lattice constant *a* of (Bi$_{1-x}$In$_x$)$_2$Se$_3$(111) films.

**Results**

**Growth and characterization of (Bi$_{1-x}$In$_x$)$_2$Se$_3$ thin films.** (Bi$_{1-x}$In$_x$)$_2$Se$_3$ films with variable thickness (2,4,7,10 and 13 QLs) were grown by MBE on single-layer graphene, prepared by graphitizing a SiC substrate[27]. For each thickness, x varies from 0% to 100%. The Bi flux was calibrated by measuring the thickness of as-grown Bi$_2$Se$_3$ films using scanning tunneling microscopy (STM) and atomic force microscopy (AFM). Then the Bi flux was fixed. The In flux was calibrated by measuring the thickness of as-grown In$_2$Se$_3$ films. In such way the flux ratio of Bi and In was determined for each x of $(Bi_{1-x}In_x)_2Se_3$, and the corresponding films were grown by simultaneously deposit Bi, In and Se atoms from independent sources. As-grown films were preliminarily characterized by reflection high-energy electron diffraction (RHEED) and STM. Figure 1a-c present typical RHEED patterns of $Bi_2Se_3$, $(Bi_{0.4}In_{0.6})_2Se_3$ and $In_2Se_3$, from which the corresponding lattice constant can be calculated. For $In_2Se_3$, the hexagonal in-plane lattice constant (*a*) is calculated as $a \approx 3.97$ Å, consistent with the value of $\alpha$-$In_2Se_3$(111) reported in previous works[28]. For $Bi_2Se_3$, $a \approx 4.13$ Å, also in agreement with the previously reported value.[29] Figure 1f summarizes the evolution of *a* with increasing chemical composition ratio *x* in $(Bi_{1-x}In_x)_2Se_3$. It is clear that *a* decreases monotonically with increasing *x*, due to the fact that In atoms are smaller than Bi atoms. Figure 1d and 1e show the STM images of as-grown $Bi_2Se_3$ and $In_2Se_3$, respectively. The height profile of the corresponding QLs can be extracted across the steps at the film edge, as shown in the insets. The QL thickness is about 0.95 nm for $Bi_2Se_3$ and 0.97 nm for $In_2Se_3$, consistent with previous works.[28] Furthermore, the sharp RHEED streaks and the atomically flat morphology demonstrate the high crystalline quality of the as-grown $(Bi_{1-x}In_x)_2Se_3$ films of variable thickness, which is suitable for further ARPES investigation.

**Dimensional crossover**

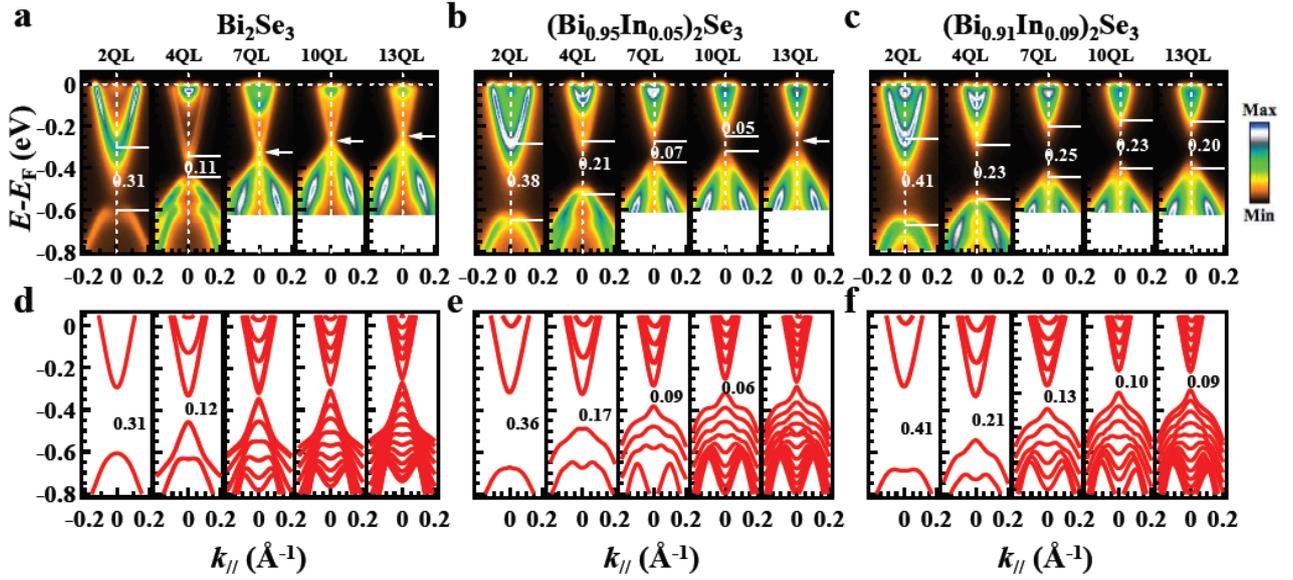

**Figure 2. Band structure evolution of $(Bi_{1-x}In_x)_2Se_3$ thin films with increasing number of layers for different substitution level x. (a,b,c),** ARPES measured band structure for different In concentration x and number of layers. The white solid lines indicate the VBM and CBM positions. The corresponding gap size (in eV) is also shown for each composition. Note they are surface state gap in (**a,b**) and bulk state gap in (**c**). The white arrows indicate the position of the Dirac point. (**d,e,f**), the corresponding theoretically calculated band structure. The size of the band gap is also indicated.

Figure 2 and Figure S1 present the band evolution of $(Bi_{1-x}In_x)_2Se_3$ films with systematic substitution and layer number control. Generally speaking, for all the spectra the Fermi level intersects the bulk conduction band due to the intrinsic doping from, *e.g.* Se vacancies. Nevertheless, we use terms like TI in the present work since we focus on the fundamental band structure rather than the transport properties that related to the position of chemical potential. We locate the position of the conduction band minimum (CBM) and valence band maximum (VBM) through energy-distribution curve (EDC) analysis (see Figure S2 and S3 for detail). The gap size, *i.e.*, the energy difference between the CBM and VBM, is shown for each substitution concentration in the corresponding panels. It is widely accepted that the surfaces of 3D TI host gapless Dirac cone in bulk crystals and thick films and this is exactly the case for $Bi_2Se_3$ films thicker than 6 QLs as shown in Figure 2a. Thus, for all the different numbers of layers, the bands of bulk and surface states can be distinguished by tracking their evolution back from the topologically nontrivial phase. In the case of x=0 (Figure 2a), as reported previously,[15] coupling between top and bottom surface state happens for $Bi_2Se_3$ films thinner than 6 QLs. Consequently, a surface state gap emerges. The size of this gap is 0.31 eV for 2 QLs and 0.11 eV for 4 QLs. When the thickness $\geq$ 7 QLs, the surface states are gapless, confirmed by the well-defined Dirac cone structure for 7 QLs, 10 QLs and 13 QLs. By using hybrid exchange-correlation functional, our DFT calculated electronic structures of $Bi_2Se_3$ films show excellent agreement with the ARPES results, especially for the band gap values (see Figure 2d). Note that theoretically there is always a small gap at the $\Gamma$ point due to the hybridization between the two surfaces, despite that the gap is too tiny and thus can be ignored within the experimental resolution. These results agree with previous works[15,19] that, with decreasing layer number, a dimensional crossover from 3D TI to 2D GSS occurs at about 6 QLs in $(Bi_{1-x}In_x)_2Se_3$ for x=0. In Figure S3 we use $2^{nd}$ curvature method[30] to treat the ARPES spectra. With such treatment the surface dispersion and gap are clearly distinguished, clearly proving the existence of gapped surface states.

With increasing In substitution x, the critical layer number corresponding to gapped-gapless surface states crossover also increases. For x=0.02 (Figure S1a and Figure S3b), the surface state gap increases to 0.35 eV and 0.17 eV for 2 QLs and 4 QLs, respectively. Moreover, a small gap about 0.03 eV opens for 7 QLs, while 10 QLs and 13 QLs remain gapless. For x=0.04 (Figure S1b and Figure S3c), the gap size further increases to 0.36, 0.19 eV and 0.05 eV for 2 QLs, 4 QLs and 7 QLs. With x increased to 0.05 (Figure 2b and Figure S3d), another surface gap of 0.05 eV opens for 10 QLs, accompanied by the continued increase of gap size for the thinner ones. This helps us to roughly locate the 3D TI-2D GSS crossover point for x=0.02 around 8 QLs, for x=0.04 around 9 QLs, and for x=0.05 around 12 QLs. The DFT calculated band structures of $(Bi_{1-x}In_x)_2Se_3$ films with x=0.05 are also in agreement with those obtained from ARPES. We note that there is discrepancy between our calculation and experimental data for thick films at x=0.09, which may attribute to the experimental non-uniform substitution of In in these samples.

The increase of dimensional crossover layer number and surface gap size with substitution level x can be qualitatively explained by the surface states penetration equation: $\lambda = h\upsilon_F / 2\pi E_g$,[31] in which $\upsilon_F$ is surface states Fermi velocity and $E_g$ is the bulk gap. Taking 6 QLs $Bi_2Se_3$ ($E_g$=0.32 eV) as a reference, its penetration depth is approximate 3 nm. From the ARPES results, $(Bi_{0.98}In_{0.02})_2Se_3$ bulk gaps are 220 meV, 205 meV and 190 meV for 7 QLs, 10 QLs and 13 QLs, respectively. According to the formula above, the surface states penetration depth can be estimated as 4.35 nm, 4.68 nm and 5.05 nm, respectively. For 7 QLs $(Bi_{0.98}In_{0.02})_2Se_3$ film, $2\lambda > 7$ nm. Consequently, coupling between top and bottom surface state happens and a narrow surface states gap emerges. For 10 QLs and 13 QLs $(Bi_{0.98}In_{0.02})_2Se_3$ films, $2\lambda < 10$ and 13 nm, so there exists still a gapless Dirac cone. For x=0.05, the bulk gaps decrease to 190meV, 160meV and 150meV, corresponding to surface penetration depth about 5.05 nm, 6 nm and 6.4 nm, for 7 QLs, 10 QLs and 13 QLs respectively. Thus, $(Bi_{0.95}In_{0.05})_2Se_3$ 7 QLs and 10 QLs are gapped while 13 QLs is gapless.

**Phase Diagram**

To approach the $(Bi_{1-x}In_x)_2Se_3$ bulk TPT point, we increase x to 9%, 17%, 20%, and 33% (Figure 2 and Figure S1). From the band structure shown for these substitution levels, no signature of Dirac cone nor surface states can be seen. We attribute the absence of the surface states to the TPT-resulted trivial topology in these films. This judgement is supported by the fact that the bulk gap size decreases with In substitution x for x≤0.05 but increases with x for x≥0.09, suggesting an inverted gap-closure and reopening process (see Figure 3a). We thus conclude that $(Bi_{1-x}In_x)_2Se_3$ films (x>9%) are in the NI phase. Furthermore, for x<0.33, the Fermi level lies in the bulk conduction band for all the thickness. When x=0.33 (see Figure 4), the bulk conduction band is almost invisible, with only a faint tail left, in agreement with the characteristic spectra reported as a variable range-hopping insulator.[22]

In topological non-trivial area, with the increasing of x, the bulk gap decreases gradually, as summarized in Figure 3a and 3b, in contrast to the sudden gap-closure scenario observed in bulk crystals.[24] Correspondingly, the surface states penetration depth increases (Figure 3c). Once the penetration depth is larger than the half of the film thickness, the top and bottom surface states begin to hybridize, leading to the gap opening in the surface states. The size of this hybridization-induced surface gap increases with substitution x, also summarized in Figure 3a. The crossover from 3D TI to 2D GSS (x=$x_H$) is thickness-dependent. The thicker the film is, the larger $x_H$ is. When x is close to the bulk critical point ($x_B$), the bulk gap vanishes, resulting in an infinite surface penetrate depth. In this case, the $(Bi_{1-x}In_x)_2Se_3$ film will open a surface state gap regardless of the thickness, which is exactly what happens in previous works.[21-24] At $x_B$, the transition from 2D GSS to NI happens, independent of the number of layers. We note here that, because of the quantum well states in thin $(Bi_{1-x}In_x)_2Se_3$ films, the negative bulk gap may not close at the bulk critical point.[19]

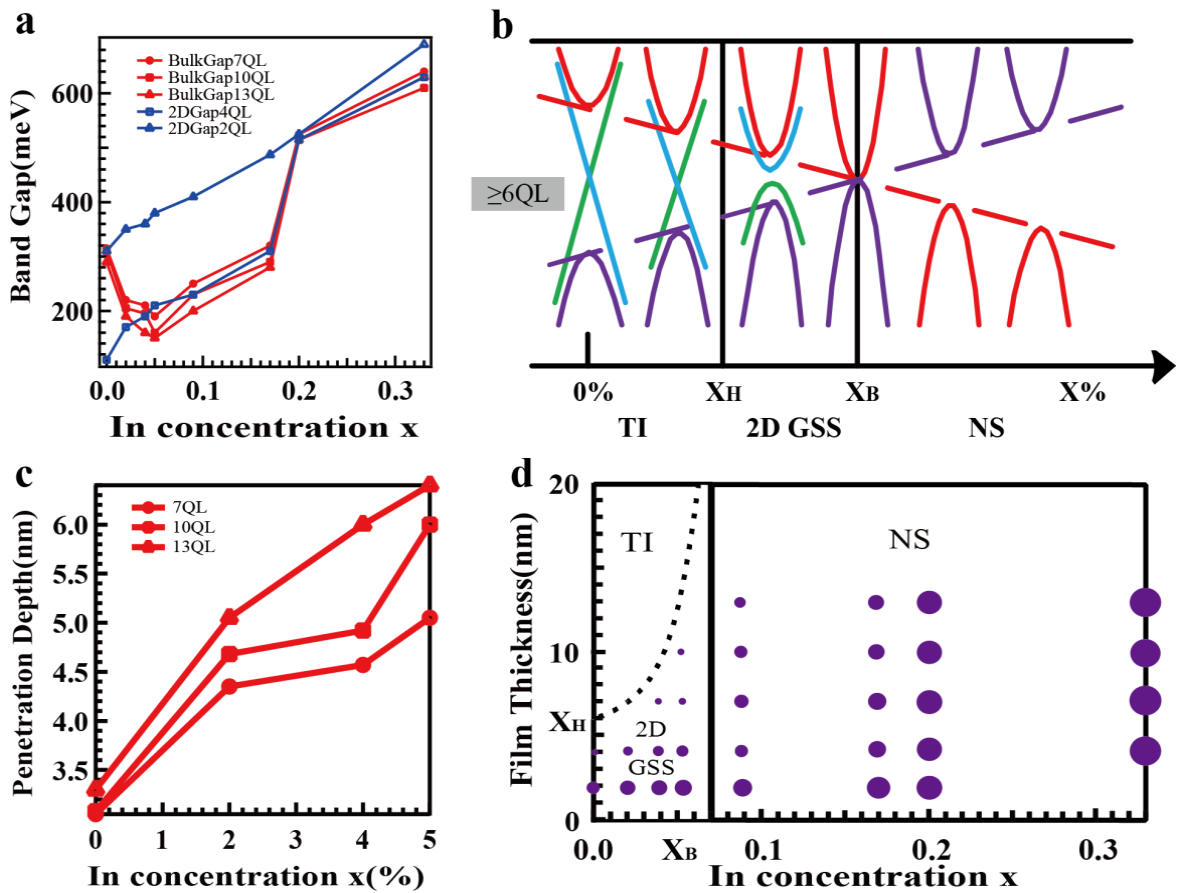

**Figure 3. $(Bi_{1-x}In_x)_2Se_3$ phase evolution with various thickness and substitution. (a)**, the 2D (in blue) and bulk (in red) gap size with various $(Bi_{1-x}In_x)_2Se_3$ In concentration for different thicknesses. **(b)**, the schematic of band structure evolution of $(Bi_{1-x}In_x)_2Se_3$ with increasing x for films thicker

than 6 QLs. **(c)**, surface states penetration depth with various $(Bi_{1-x}In_x)_2Se_3$ thickness and x. **(d)**, the $(Bi_{1-x}In_x)_2Se_3$ phase diagram with various film thickness and x, the solid circles represent all the data points from our ARPES results, with their size proportional to the gap size.

Figure 3d demonstrates a phase diagram with the variable of x and film thickness concluded from our ARPES results. From this diagram, as indicated by the black dashed line, the dimensional crossover from 3D TI to 2D GSS state ($x=x_H$) is thickness dependent and approaches the bulk limit $x_B$ asymptotically. This naturally provides an experimental way to tune the dimensional crossover in $(Bi_{1-x}In_x)_2Se_3$ films by band gap engineering through thickness and substitution control. This phase evolution manipulation, together with the ability to grow atomically flat ultrathin films also proved in this work, would pave the route for the exploration of novel 2D topological states and potential device application.

**Topological nature of thin films**

Our ARPES results not only show the bulk TPT and the dimensional crossover of $(Bi_{1-x}In_x)_2Se_3$ films (see Figure 3), but also reveal the topological feature of $Bi_2Se_3$ thin films as 2D GSS systems. In principle, we can create a dynamical evolution process and see if TPT happens during the process by introducing an external knob, such as the strength of SOC, electric field or strain. By tuning the knob, we turn the initial state with unknown topological feature to a known final state, say, a NI state. If the initial state is topologically nontrivial, we would expect the band gap decreases to zero and reopens during the evolution; otherwise the initial state is topologically trivial. Here we consider such In substitution as an external knob that mainly modulates the strength of SOC of the system, while the lattice constant also decreases with doping. Figure 3a and 4a shows the gap evolution for the case of 4 QLs as a function of In concentration x. The band gaps exhibit a monotonic increasing trend upon doping till x=0.20, far beyond the bulk critical point $x_B$, indicating that 4 QLs $Bi_2Se_3$ is a 2D NI rather than 2D quantum spin Hall insulator. Theoretically, $Z_2$ invariant of $Bi_2Se_3$ QLs can be computed by Fu-Kane formula,[32] which counts the product of the parities of all occupied states and all time-reversal-invariant momenta (TRIM). Our DFT calculations show that the thin films $Bi_2Se_3$ with an observable hybridization gap (i.e., 1-5 QLs) are all topologically trivial, in contrast to the previously reported oscillatory behaviour between NI and TI from 3 QLs.[16,17]

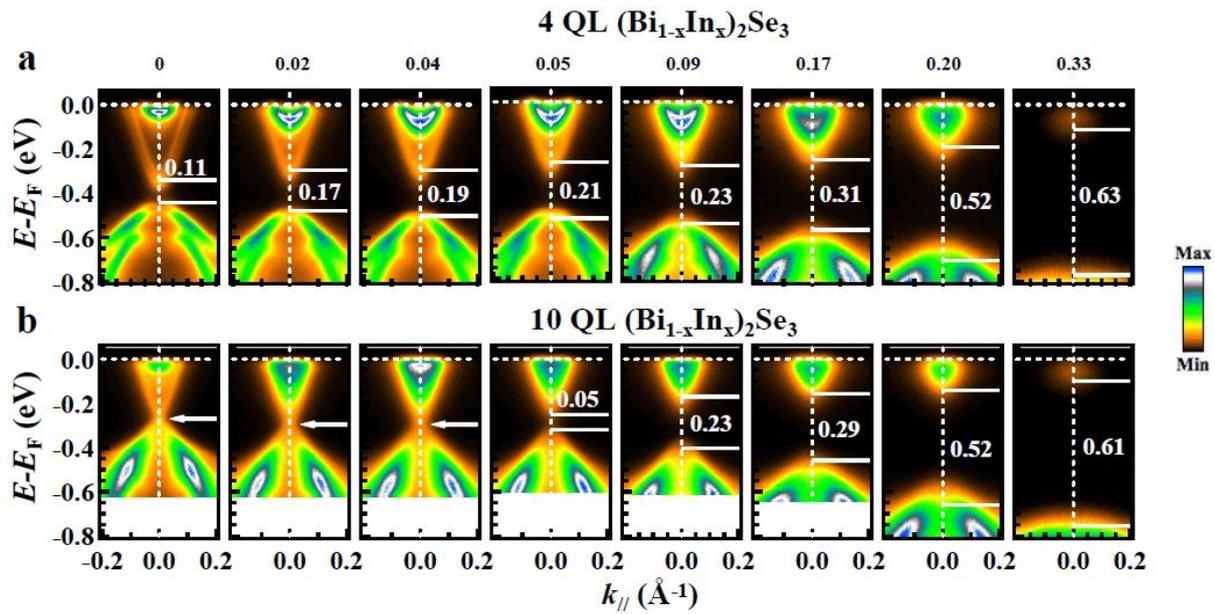

**Figure 4: Band structure evolution of $(Bi_{1-x}In_x)_2Se_3$ thin films for 4 QLs and 10 QLs with increasing substitution level x.** The white solid lines indicate the VBM and CBM positions. The corresponding gap size (in eV) is also shown for each composition. The white arrows indicate the position of the Dirac point.

As shown in Figure 4b and 3a, for 10 QLs, the gap evolution has similar trend compared with those of 4 QLs for x≥0.05. On the other hand, although theoretically there is always a gap at Dirac cone due to the hybridization between the opposite surfaces in a thick film, the initial states of 7 QLs and 10 QLs films show "gapless" surface dispersion seen by ARPES as the thickness is larger than the twice of surface penetration depth. Consequently, for films thicker than 6 QLs, a dimensional crossover in topological nontrivial phase occurs before the TPT happens at $x_B$. Therefore, it is not quite meaningful to treat these thick films as 2D gapped states and judge if the tiny gap is inverted or not.

**Conclusion**

Our work represents a systematic study of a classic 3D TI system in its ultrathin form with precise thickness and substitution control. The ARPES spectra clearly describe how the electronic structure evolves when both SOC and surface state hybridization come into play, making an essential step towards controlling topological phase evolution for fundamental research, material engineering and device fabrication. The combining ARPES and DFT band gap evolution analysis also demonstrates a practical solution to determine the topology of one particular phase through its phase evolution to another phase with known topology.

In the case of device fabrication for topology-based transport, the current conductive substrate, graphene on SiC, must be replaced for insulating substrate such as hexagonal boron nitride. Further experimental efforts are needed to optimize the growth parameters and explore potential issues such as the interfacial interaction between TI films and the substrate.

**Methods**

**Sample preparation.** We use SiC epitaxy 1~2 layers graphene as the substrate for $(Bi_{1-x}In_x)_2Se_3$. The SiC substrates were first degassed by heating to 600℃ for 2 hours in the MBE chamber until the system returns to the base pressure. Then we heat SiC to 900℃ to remove the surface oxides. During this period, Si is evaporated to decrease the loss speed of Si in SiC. At last, a flat single layer graphene was formed by heating SiC substrates to 1300℃ for 10 minutes. The Bi, Se and In sources were thermally evaporated from Knudsen cells at the same time with Bi and Se fixed on 420℃ and 130℃. In concentration x was controlled by changing In evaporating temperature. The graphene substrate was kept at 200℃ while growing.

**Density functional theory.** We first calculate the electronic structures of bulk $(Bi_{1-x}In_x)_2Se_3$ alloys by first-principles calculations using the projector-augmented wave (PAW) pseudopotentials,[33] as implemented in the Vienna Simulation Package (VASP).[34] For the exchange-correlation energy, we used the screened hybrid density functional of the Heyd–Scuseria–Ernzerhof type (HSE06).[35] A 6×6×6 K-mesh is selected for the sampling of the Brillouin zone. The energy cutoff is set to 400 eV, while the total energy minimization is performed with a tolerance of $10^{-6}$ eV. The ions are relaxed until the atomic force is less than $10^{-3}$ eV/Å with fixed lattice constants a= 4.13 Å, c=28.58 Å which are consistent with the previously reported value.[28] Spin-orbit coupling is included in throughout the calculations self-consistently. We simulate the composition alloying effects by using virtual crystal approximation (VCA) method,[36] which considers a crystal composed of fictitious "virtual" atoms that interpolate between the behavior of the atoms in the parent compounds. In our case, the potential of $(Bi_{1-x}In_x)_2Se_3$ alloy is generated by compositionally averaging the pseudopotentials of Bi and In, i.e., $V_x = xV_{In} + (1-x)V_{Bi}$. We constructed Wannier representations by projecting the Bloch states from the first-principles calculations of bulk materials onto the $p$ orbitals of the cations and anions. The band structures of the slabs with different quintuple layers are calculated in the tight-binding models constructed by these Wannier representations,[37-39] as implemented in the WannierTools package.[40]

ASSOCIATED CONTENT

**Supporting Information**.

Band structure evolution of $(Bi_{1-x}In_x)_2Se_3$ thin films with increasing number of layers for different substitution level x, band structure of 20 QLs $(Bi_{0.91}In_{0.09})_2Se_3$ thin film and the method to determine the bulk gap size, band structure spectra of $(Bi_{1-x}In_x)_2Se_3$ (x=0, 0.02, 0.04, 0.05) thin films which treated with 2nd curvature method to highlight the surface state dispersion, band structure of $In_2Se_3$ thin films with different thickness. Supporting information is available free of charge.

AUTHOR INFORMATION

**Corresponding Author**

*E-mail: oscarwang2008@sina.com; liuqh@sustech.edu.cn; chency@sustech.edu.cn.

**Author Contributions**

The manuscript was written through contributions of all authors. All authors have given approval to the final version of the manuscript. †These authors contributed equally to this paper.

**Notes**

The authors declare no competing financial interest.

ACKNOWLEDGMENT

We thank Prof. Hai-Zhou Lu for helpful discussions. This work was supported by the National Natural Science Foundation of China, the Ministry of Science and Technology of China, the General Program of Beijing Academy of Quantum Information Sciences (Project No.Y18G17), the Opening Foundation of State Key Laboratory of High Performance Computing (201601-02), Science, Technology and Innovation Commission of Shenzhen Municipality (No. ZDSYS20170303165926217, No. JCYJ20170412152620376).

open-source software package for novel topological materials. *Comput. Phys. Commun.* **2018**, 224, 405.